\documentclass[reprint,showpacs,twocolumn,superscriptaddress,fleqn,floatfix]{revtex4-1}
\pdfoutput=1
\usepackage{graphics}
\usepackage{graphicx}
\usepackage{amsmath,amssymb,amsthm,amsfonts,epsfig}
\usepackage{gensymb}
\usepackage{bm}
\usepackage{natbib}
\usepackage[version=3]{mhchem}
\usepackage{hyperref}
\usepackage{lipsum}
\usepackage{color}
\usepackage{gensymb}
\usepackage{longtable,afterpage}
\usepackage{lipsum}
\hypersetup{%
  breaklinks=true,%
  colorlinks=true,%
  citecolor=blue,%
  filecolor=blue,%
  linkcolor=blue,%
  urlcolor=blue%
}
\newcommand{\figref}[1]{Fig.~\ref{#1}}
\newcommand{\tabref}[1]{Table~\ref{#1}}
\begin{document}
%
%
\title{Boron phosphide as a \emph{p}-type transparent conductor: optical absorption and transport through electron-phonon coupling}
%
\author{Viet-Anh Ha}
\email[Present address: Oden Institute for Computational Engineering and Sciences, University of Texas at Austin, 201 E. 24$^{th}$ Street, Austin, TX 78712, USA]{}
 \affiliation{Institute of Condensed Matter and Nanoscience (IMCN), Universit\'{e} Catholique de Louvain (UCL), Chemin \'{e}toiles 8, bte L7.03.01, Louvain-la-Neuve 1348, Belgium}
\author{Bora Karasulu}%
 \affiliation{Cavendish Laboratory, University of Cambridge, J. J. Thomson Avenue, Cambridge CB3 0HE, United Kingdom}%
 \author{Ryo Maezono}%
 \affiliation{School of Information Science, JAIST, Asahidai 1-1, Nomi, Ishikawa 923-1292, Japan}%
 \author{Guillaume Brunin}
 \affiliation{Institute of Condensed Matter and Nanoscience (IMCN), Universit\'{e} Catholique de Louvain (UCL), Chemin \'{e}toiles 8, bte L7.03.01, Louvain-la-Neuve 1348, Belgium}
 \author{Joel Basile Varley}%
 \affiliation{Lawrence Livermore National Laboratory 7000 East Avenue, L-413 Livermore, CA 94550, USA}%
\author{Gian-Marco Rignanese}%
 \affiliation{Institute of Condensed Matter and Nanoscience (IMCN), Universit\'{e} Catholique de Louvain (UCL), Chemin \'{e}toiles 8, bte L7.03.01, Louvain-la-Neuve 1348, Belgium}%
\author{Bartomeu Monserrat}%
 \email[\emph{E-mail}: ]{bm418@cam.ac.uk}
 \affiliation{Cavendish Laboratory, University of Cambridge, J. J. Thomson Avenue, Cambridge CB3 0HE, United Kingdom}
 \affiliation{Department of Materials Science and Metallurgy, University of Cambridge, 27 Charles Babbage Road, Cambridge CB3 0FS, United Kingdom}%
\author{Geoffroy Hautier}%
 \email[\emph{E-mail}: ]{geoffroy.hautier@uclouvain.be}
 \affiliation{Institute of Condensed Matter and Nanoscience (IMCN), Universit\'{e} Catholique de Louvain (UCL), Chemin \'{e}toiles 8, bte L7.03.01, Louvain-la-Neuve 1348, Belgium}%
\date{\today}

\begin{abstract}
Boron phosphide has recently been identified as a potential high hole mobility transparent conducting material. This promise arises from its low hole effective masses. However, BP has a relatively small 2 eV indirect band gap which will affect its transparency. In this work, we computationally study both optical absorption across the indirect gap and phonon-limited electronic transport to quantify the potential of boron phosphide as a $p$-type transparent conductor. We find that phonon-mediated indirect optical absorption is weak in the visible spectrum and that the phonon-limited hole mobility is very high (around 900 cm$^2$/Vs) at room temperature. This exceptional mobility comes from a combination of low hole effective mass and very weak scattering by polar phonon modes. We rationalize the weak scattering by the less ionic bonding in boron phosphide compared to oxides. We suggest this could be a general advantage of non-oxides for $p$-type transparent conducting applications. Using our computed properties, we assess the transparent conductor figure of merit of boron phosphide and shows that it exceeds by one order of magnitude that of established $p$-type transparent conductors, confirming the potential of this material. 
%
\end{abstract}

\maketitle


\section{\label{intro}Introduction}

Transparent conducting materials (TCMs) are necessary to many applications ranging from solar cells to transparent electronics. So far, \emph{n}-type oxides (e.g., \ce{In2O3}, \ce{SnO2} and \ce{ZnO}) are the highest performing TCMs, allowing them to be used in commercial devices~\cite{H.Ohta2004, A.Facchetti2010, K.Ellmer2012, P.Barquinha2012, E.Fortunato2012}. On the other hand, \emph{p}-type TCMs show poorer performances, especially in terms of carrier mobility, limiting the development of new technologies such as transparent solar cells or transistors~\cite{K.Ellmer2012, S.C.Dixon2016}. Hence, the search for high performance \emph{p}-type TCMs has been a long-lasting goal of the materials research community. 

As demonstrated by analyzing high-throughput computational data, \emph{p}-type oxides have inherently higher effective masses than \emph{n}-type oxides, thus rationalizing the current gap in mobility between the best \emph{p}-type and \emph{n}-type oxides~\cite{G.Hautier2013, G.Brunin2019b}. The strong oxygen \emph{p}-orbital character in the valence band of most oxides is responsible for their statistically high hole effective mass. This inherent difficulty in developing \emph{p}-type transparent oxides with low hole effective mass justifies moving towards non-oxide TCM chemistries, including fluorides~\cite{H.Yanagi2003}, sulfides~\cite{S.Park2002, R.W.Robinson2016}, oxyanions~\cite{K.Ueda2000, Williamson2019}, suboxides~\cite{J.B.Varley2014} or germanides~\cite{F.Yan2015}. The opportunities in non-oxide chemistries have been recently confirmed by further analysis of high-throughput computational data showing that non-oxide materials have statistically lower hole effective masses than oxides~\cite{J.B.Varley2017,V.Ha2019}. Unfortunately, these lower hole effective masses come with smaller band gaps that are detrimental to transparency. Using the difference between fundamental (indirect) and optical (direct) band gaps, we identified through high-throughput computing boron phosphide (BP) as a very promising \emph{p}-type TCM candidate~\cite{J.B.Varley2017}. Boron phosphide shows according to computations a rare combination of low hole effective mass, large direct band gap, and \emph{p}-type dopability. 

Boron phosphide was characterized experimentally through electrical and optical measurements ~\cite{K.Shohno1974, M.Takigawa1974, Y.Kumashiro1988, Y.Kumashiro1989, Y.Kumashiro1990, Y.Kumashiro2010, Stone1960, M.Iwami1975, M.Odawara2005} but the sample quality is variable between these studies and an in-depth theoretical analysis of boron phosphide is thus required. In this paper, we theoretically study indirect optical absorption and transport properties of hole carriers in BP. Using state-of-the-art electron-phonon computations, we investigate how phonon-assisted indirect optical transitions impact the transparency of this material. We also use the electron-phonon scattering matrix elements to study transport properties and especially hole mobility. We combine these theoretical results to assess the performance of boron phosphide in terms of hole conductivity and transparency, evaluating its transparent conducting material figure of merit (FOM). Our results show that BP can outperform current \emph{p}-type transparent conducting oxides.

\section{\label{method}Computational details}

\textbf{Indirect optical absorption}
The complex dielectric function $\varepsilon_1(E)+i\varepsilon_2(E)$ determines the extinction coefficient $\kappa=[(-\varepsilon_1+\sqrt{\varepsilon_1^2+\varepsilon_2^2})/2]^{1/2}$ and the absorption coefficient $\alpha(E)=2E\kappa(E)/\hbar c$~\cite{AmbroschDraxl2006}, where $E$ is the photon energy, $\hbar$ is the reduced Planck constant and $c$ is the speed of light. In the independent-particle approximation and the electric dipole approximation, the frequency-dependent imaginary part of the dielectric function is given by
\begin{equation}
\begin{split}
\varepsilon_2(E)=\frac{2\pi}{mN}\frac{\hbar^2\omega^2_{\mathrm{p}}}{E^2}\sum_{v,c} \int_{\mathrm{BZ}} & \frac{d\mathbf{k}}{(2\pi)^3} |M_{cv\mathbf{k}}|^2 & \\ & \times \delta(\epsilon_{c\mathbf{k}}-\epsilon_{v\mathbf{k}}-E),
\end{split}
\label{dielectric}
\end{equation}
where $m$ is the electron mass, $N$ is the number of electrons per unit of volume,  $\omega^2_{\mathrm{p}}=Ne^2/\epsilon_0 m$ is the plasma frequency of the solid~\cite{F.Giustino2014} with $e$ the elementary charge and $\epsilon_0$ the vacuum permittivity, $M_{cv\mathbf{k}}=\langle\psi_{c\mathbf{k}}|\hat{\mathbf{e}}\cdot\mathbf{p}|\psi_{v\mathbf{k}}\rangle$ is the optical matrix element where $\hat{\mathbf{e}}$ is the polarization of the incident light and $\mathbf{p}$ is the momentum operator, and electronic wave functions $|\psi\rangle$ of energy $\epsilon$ and momentum $\mathbf{k}$ are labelled by their valence $v$ or conduction $c$ band index. The factor 2 in \eqref{dielectric} represents for the spin degeneracy. The Kramers-Kronig relation~\cite{Ziman1972} gives the real part of the dielectric function $\varepsilon_1(E)$.

The effects of lattice dynamics on the dielectric function at temperature $T$ are introduced by means of the Williams-Lax theory~\cite{Williams1951,Lax1952} as
\begin{equation}
\begin{split}
\varepsilon_2(E;T)=\frac{1}{\mathcal{Z}}\sum_{\mathbf{s}}\langle\Phi_{\mathbf{s}}(\mathbf{u})| \varepsilon_2(E;\mathbf{u})|\Phi_{\mathbf{s}}(\mathbf{u})\rangle & \\
\times e^{-E_{\mathbf{s}}/k_{\mathrm{B}}T},
\end{split}
\label{dielectric_temperature}
\end{equation}
where $|\Phi_{\mathbf{s}}(\mathbf{u})\rangle$ is the harmonic vibrational wave function in state $\mathbf{s}$ with energy $E_{\mathbf{s}}$, $\mathbf{u}=\{u_{\nu\mathbf{q}}\}$ is a collective ionic coordinate in terms of normal modes of vibration $(\nu,\mathbf{q})$, $\mathcal{Z}=\sum_{\mathbf{s}}e^{-E_{\mathbf{s}}/k_{\mathrm{B}}T}$ is the partition function, and $k_{\mathrm{B}}$ is Boltzmann's constant. This expression has recently become amenable to first-principles methods~\cite{Patrick2014,Zacharias2015,Zacharias2016,Monserrat2018_bso,Kang2018,Morris2018,Bravic2019} and we evaluate it using thermal lines~\cite{Monserrat2016_TL,Monserrat2018_JPCM}.

We perform the optical absorption calculations at finite temperature using DFT in the projector augmented-wave formulation as implemented in the {\sc vasp} package~\cite{vasp1,vasp2,vasp3,vasp4}. We perform self-consistent and lattice dynamics calculations using an energy cutoff of $500$~eV and an electronic Brillouin zone (BZ) Monkhorst-Pack~\cite{MonkhorstPack} sampling grid of size $8\times8\times8$ for the primitive cell and commensurate grids for the supercells. We perform calculations using both the generalized gradient approximation of Perdew-Burke-Ernzerhof (PBE) \cite{PBE} and the hybrid Heyd-Scuseria-Ernzerhof (HSE) \cite{HSE06} functionals. We calculate the harmonic lattice dynamics using the finite displacement method in conjunction with nondiagonal supercells~\cite{LloydWilliams2015} using coarse vibrational BZ grids of sizes up to $4\times4\times4$.\\

\textbf{Transport properties}
From Drude's theory~\cite{charleskittel2004}, the mobility, written as $\mu = e\tau/m^{*}$, is proportional to the average relaxation time $\tau$ and inversely proportional to the effective mass $m^{*}$ of carriers. The relaxation time $\tau$ (inverse of scattering rate) depends on different scattering mechanisms such as the scattering by phonons, ionized and neutral impurities, or grain boundaries. Here, we only take into account the scattering of carriers by phonons, which is likely to be an important component of scattering and is an intrinsic mechanism. The carrier scattering by phonons can be computed theoretically if the electron-phonon coupling matrix elements are known. In principle, one can employ density-functional perturbation theory (DFPT) to obtain the electron-phonon matrix elements from first principles~\cite{PhysRevLett.58.1861, PhysRevA.52.1086, Giustino2017}. However, convergence of relevant physical properties (e.g. electron scattering rate by phonons) often requires very dense \textbf{k}- and \textbf{q}-point meshes for electrons and phonons respectively and a considerable computational time if fully performed within DFPT. The recently developed interpolation techniques based on Wannier functions offer a very practical and efficient solution to overcome this obstacle. Here, we use the EPW code~\cite{J.Noffsinger2010, S.Ponce2016} interfaced with Quantum ESPRESSO (QE)~\cite{P.Giannozzi2009, P.Giannozzi2017} to calculate the relaxation time $\tau_{n\textbf{k}}$ (\emph{n} is the band index). More details on the theory and the implementation are presented in Ref.~\cite{S.Ponce2016}. In this work, the electron-phonon interaction matrix elements were computed with QE using DFPT on a coarse $6\times6\times6$ \textbf{q}-point mesh as a starting point for the interpolation with EPW. $\tau_{n\textbf{k}}$ is calculated using EPW on dense $80\times80\times80$ meshes for both \textbf{k}-point (for electrons) and \textbf{q}-point (for phonons) to guarantee the convergence. The Fermi level is set correspondingly to each doping concentration. All phonon modes for both inter-band and intra-band scattering mechanisms are taken into account in the computation of scattering rates. The structure relaxation, self-consistent, non self-consistent and phonon calculations are performed using the PBE functional and norm-conserving pseudopotentials~\cite{PhysRevLett.43.1494} with very stringent parameters for convergence, e.g. a high cut-off energy of 1088.5~eV ($80$~Ry).

In order to estimate the hole mobility due to the phonon scattering, we solve Boltzmann's transport equation (BTE) using the BoltzTrap package~\cite{G.K.H.Madsen2006}. The relaxation time $\tau_{n\textbf{k}}$ obtained with EPW and DFT band structure $\epsilon_{n\textbf{k}}$ are necessary inputs for BoltzTrap. More details about the calculations can be found in Ref.~\cite{V.Ha2019}\\

\textbf{Figure of merit} 
In many applications, the quantities of interest for TCMs are the conductivity and transparency. It is convenient to use figures of merits (FOMs) that will estimate the performance of a TCM material through one quantity. Different FOMs exist in the literature to compare TCMs~\cite{K.Ellmer2012}. We use the one defined by Haacke~\cite{Haacke1976} as $\overline{T}^{10}\sigma_s$, where $\overline{T}$ is the transparency or transmittance and $\sigma_s = \sigma t$ is the sheet conductivity of a film with thickness $t$ and conductivity $\sigma$. The conductivity is simply given by $\sigma = \mu C e$ with $\mu$ and $C$ being the hole mobility and concentration, respectively.
The average transmittance in the visible spectrum is computed as
\begin{equation}
\overline{T} = \frac{\int_{\mathrm{vis}} B_E [1-R(E)] \exp[-\alpha(E)t] dE}{\int_{\mathrm{vis}} B_E dE},
\end{equation}
where $B_E$ is the spectral radiance of a black body at a temperature of $5778$ K to model the solar spectrum, $R(E)$ is the reflectivity, and $t$ is the thickness of a given film. $\alpha(E)$ is computed as the sum of the indirect absorption (the direct absorption is negligible in the visible spectrum as the direct gap of BP $>$ 3~eV~\cite{J.B.Varley2017}) and the absorption due to plasmon effects which is modeled through the Drude model as in Ref.~\cite{G.Brunin2019}. The reflectivity is derived from the absorption coefficient and the refractive index, obtained through Kramers-Kronig relations~\cite{Ziman1972}.
We compare the FOM of BP with that of a current \emph{p}-type TCM, \ce{CuAlO2}. For BP, the theoretical mobility and absorption coefficient are used. For \ce{CuAlO2}, we use a hole mobility of 1~cm$^2$/Vs~\cite{H.Kawazoe97, H.Yanagi2000, R.-S.Yu07, J.Tate09} and an effective mass of $2.5$ times the free electron mass~\cite{G.Hautier2013} as inputs for the Drude model to obtain the transmittance and the conductivity, as in Ref.~\cite{G.Brunin2019}. In this case, our calculation does not rely on the absorption coefficient but rather on a relative dielectric constant of 11.7 in order to model the 70\% transmittance measured in thin films~\cite{H.Yanagi2000}. We neglect the influence of the second gap, which has been theoretically computed for \ce{CuAlO2}~\cite{V.-A.Ha16}. For both materials, we consider films of varying thicknesses and hole concentrations.

\section{\label{results}Results and Discussion}
\begin{figure}[!htb]
\begin{center}
\includegraphics[width=0.9\linewidth]{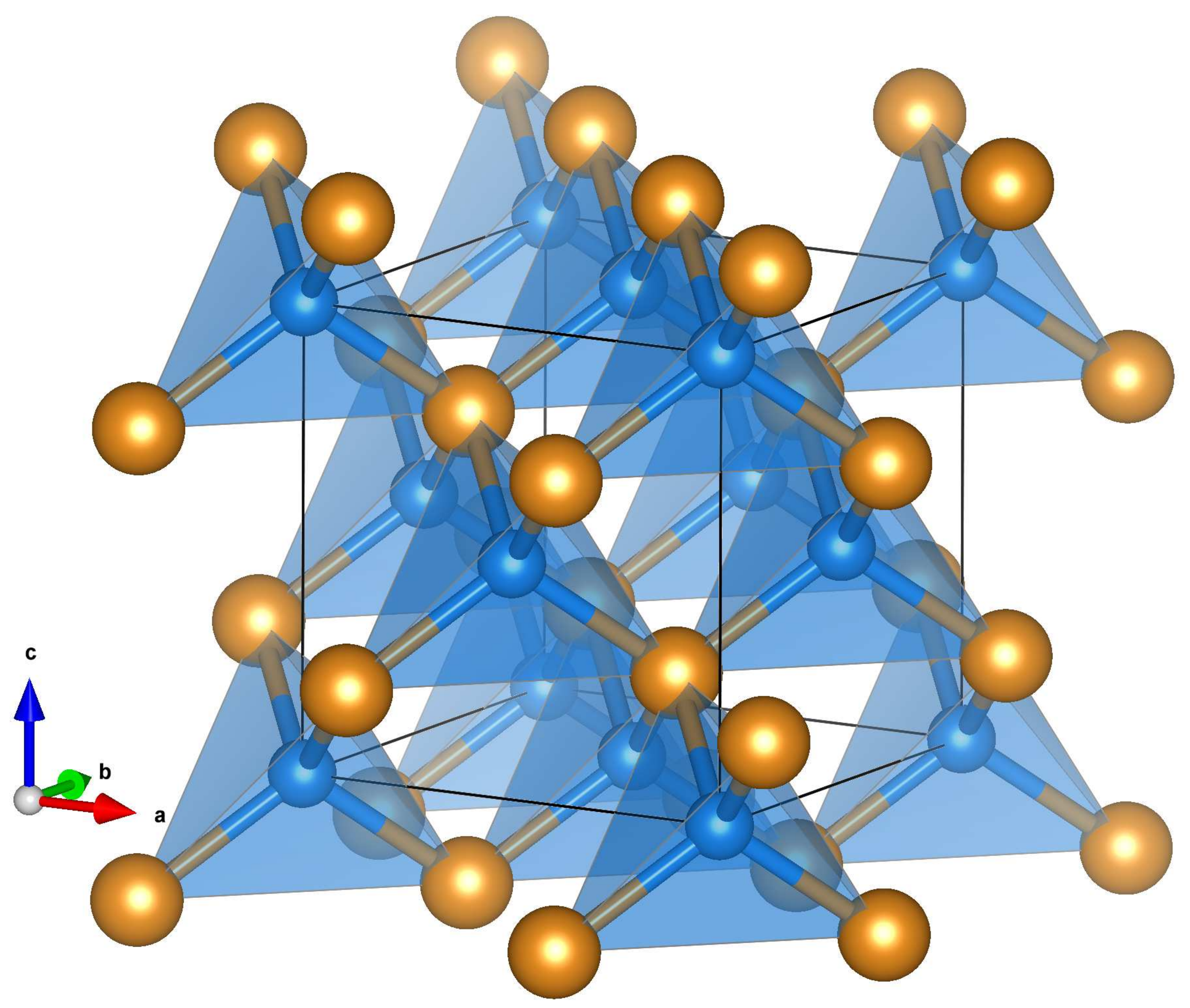}
\end{center}
\vspace{-15pt}
\caption{The conventional cell of BP with tetrahedral local environments around B atoms (blue).}
\label{concell}
\end{figure}

\begin{figure}[!htb]
\begin{center}
\includegraphics[width=0.9\linewidth]{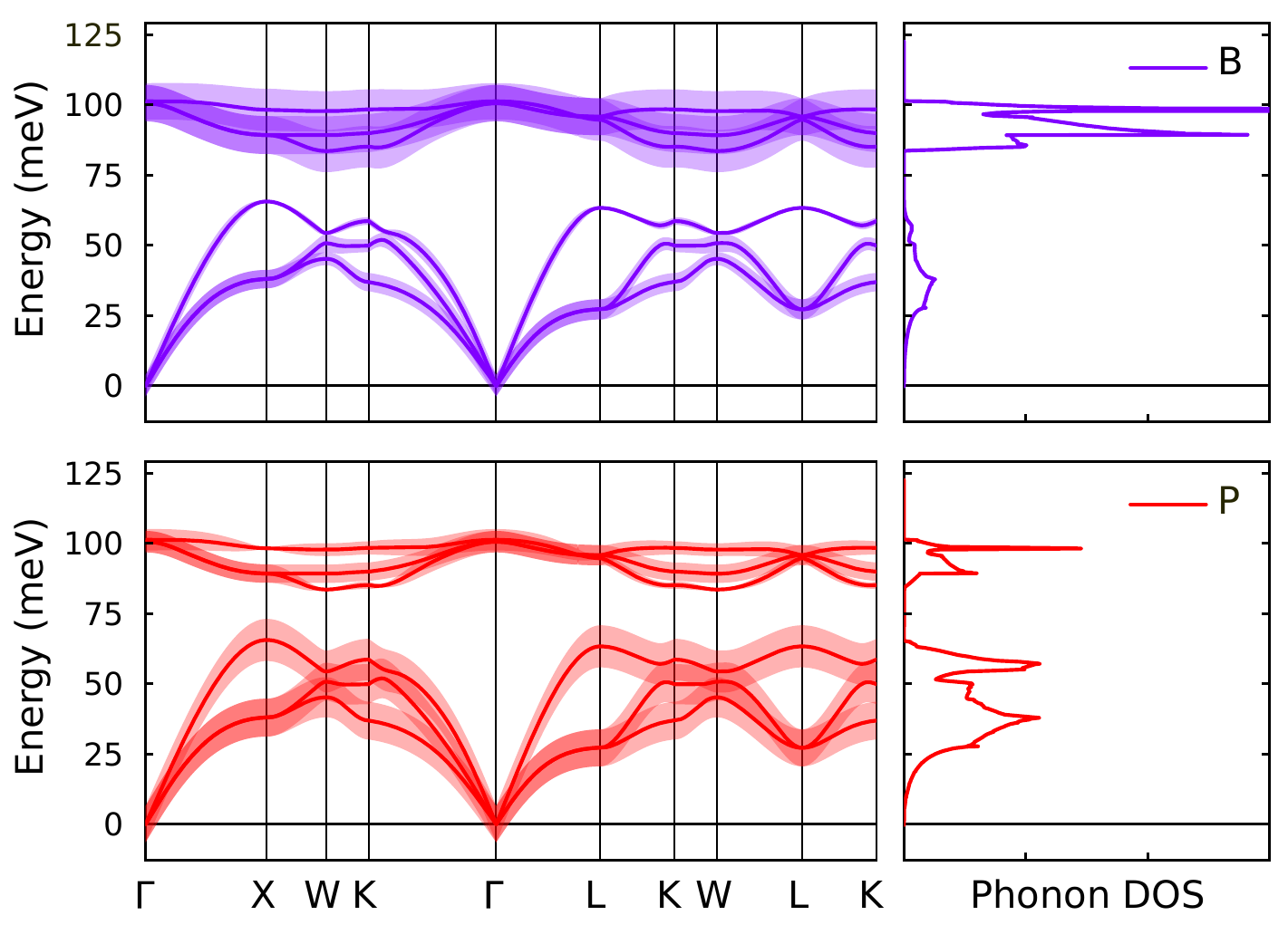}
\end{center}
\vspace{-15pt}
\caption{Phonon dispersions with fat bands representing atomic displacements associated with lattice vibrations. The width of fat bands gives a qualitative understanding of the what species are involved in the phonon modes. The projected DOS of phonons on each type of atom are correspondingly shown next to the phonon dispersion. Here, the calculations are done using semilocal PBE functional.}
\label{phband}
\end{figure}

The conventional cell of BP is shown in \figref{concell}. Each B atom is surrounded by four P atoms in tetrahedral corner-sharing local environments. The cubic symmetry leads to an isotropic effective mass tensor~\cite{J.B.Varley2017}. \figref{phband} shows the phonon dispersions (fat bands) and projected density of states (DOS) of phonons for BP using the PBE functional. The fat bands representation indicates which atom participates in the phonon modes. The lighter B atoms mainly contribute to the optical modes at high frequencies (3 modes) while the heavier P atoms play an important role in the three acoustic modes at low frequencies. The phonon dispersion computed using the HSE functional leads to a very similar dispersion with slightly higher phonon frequencies, particularly for the optical modes (see Fig. S1 of the supplementary document).

%

%
%

\begin{figure}[!htb]
\begin{center}
\includegraphics[width=0.9\linewidth]{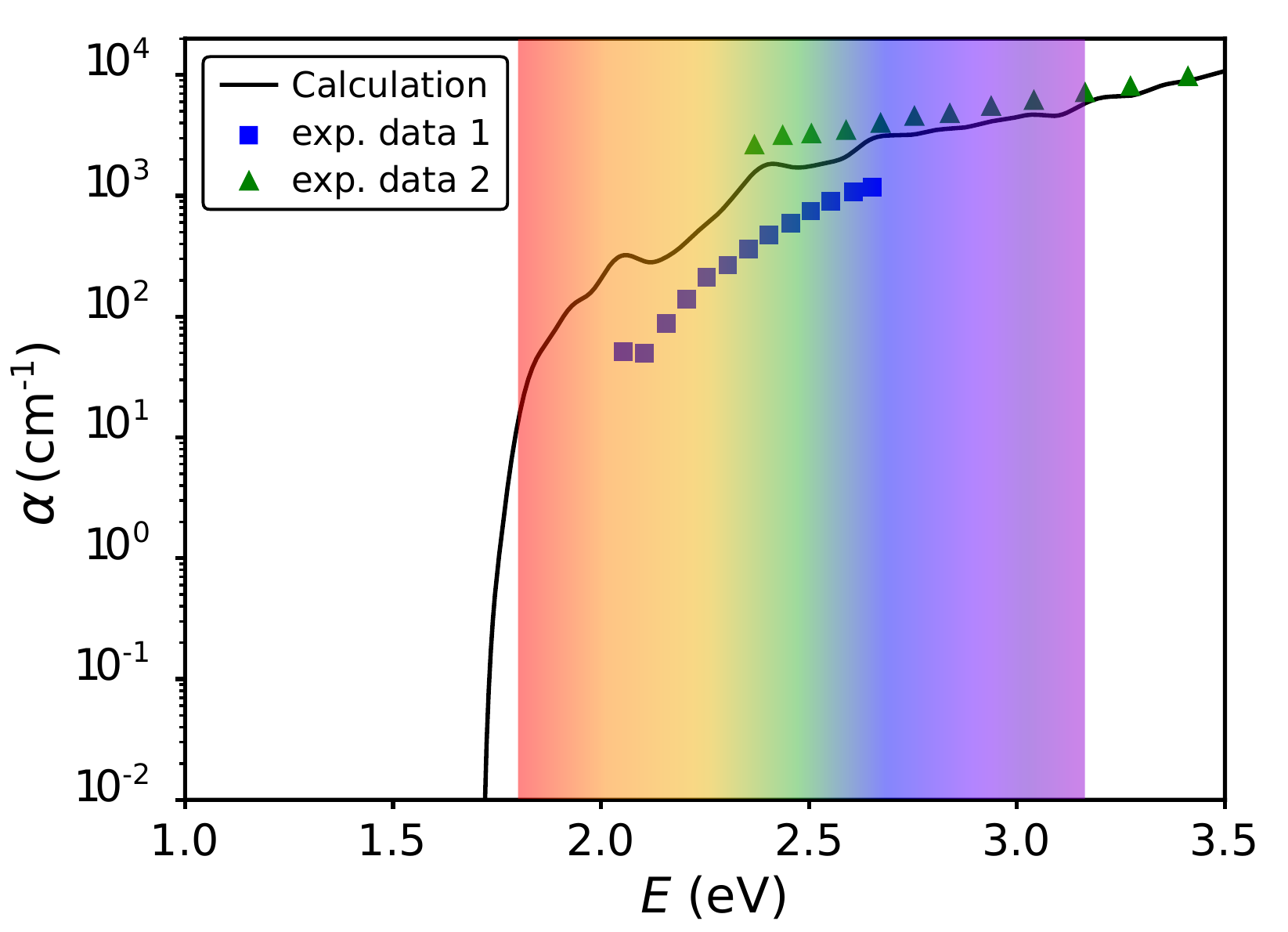}
\end{center}
\vspace{-15pt}
\caption{Indirect absorption coefficient as a function of photon energy calculated using the HSE functional at $300$~K. The experimental data from Refs.~\cite{M.Iwami1975} (exp. data 1) and \cite{M.Odawara2005} (exp. data 2) are also replotted for comparison. The colour band indicates the visible spectrum.}
\label{indabs}
\end{figure}

The phonon-assisted optical absorption at $300$~K, $\alpha(E)$, is shown in \figref{indabs}, which is calculated using the HSE functional for both electronic band structure and phonon dispersion. The indirect band gap has a static lattice (in an assumption that nuclei do not vibrate around their equilibrium positions) value of 1.98~eV, while the direct optical gap is about 4.34~eV and therefore the energy range shown in \figref{indabs} only corresponds to indirect absorption. As expected for indirect transitions, $\alpha(E)$ is weak below the direct band gap with an average value of about $10^3$ cm$^{-1}$ in the visible range. The absorption onset at $300$~K is red-shifted by about 0.25~eV in comparison with the indirect HSE static gap of 1.98~eV because the temperature dependence of the electronic band structure is also included in our calculations. The same observation was recently reported for one of the best \emph{n}-type TCOs, \ce{BaSnO3}~\cite{Monserrat2018_bso}. We also include a comparison with experimental data from Refs.~\cite{M.Iwami1975, M.Odawara2005} in the figure, which shows reasonable agreement with our calculations. The remaining small differences might come from the approximation of the \emph{ab initio} computational framework but might also very likely originate from issues with sample quality in the experimental reports. It is worth noting that this computational framework was utilized to compute indirect absorption for bulk silicon yielding excellent agreement with experiment~\cite{Zacharias2015}. We note that analogous calculations using the semilocal PBE functional lead to a qualitatively similar absorption profile, but with a significant red shift of about $1$~eV associated with the standard band gap underestimation of such semilocal functional.

%

\begin{figure*}[!htb]
\begin{center}
\includegraphics[width=0.95\linewidth]{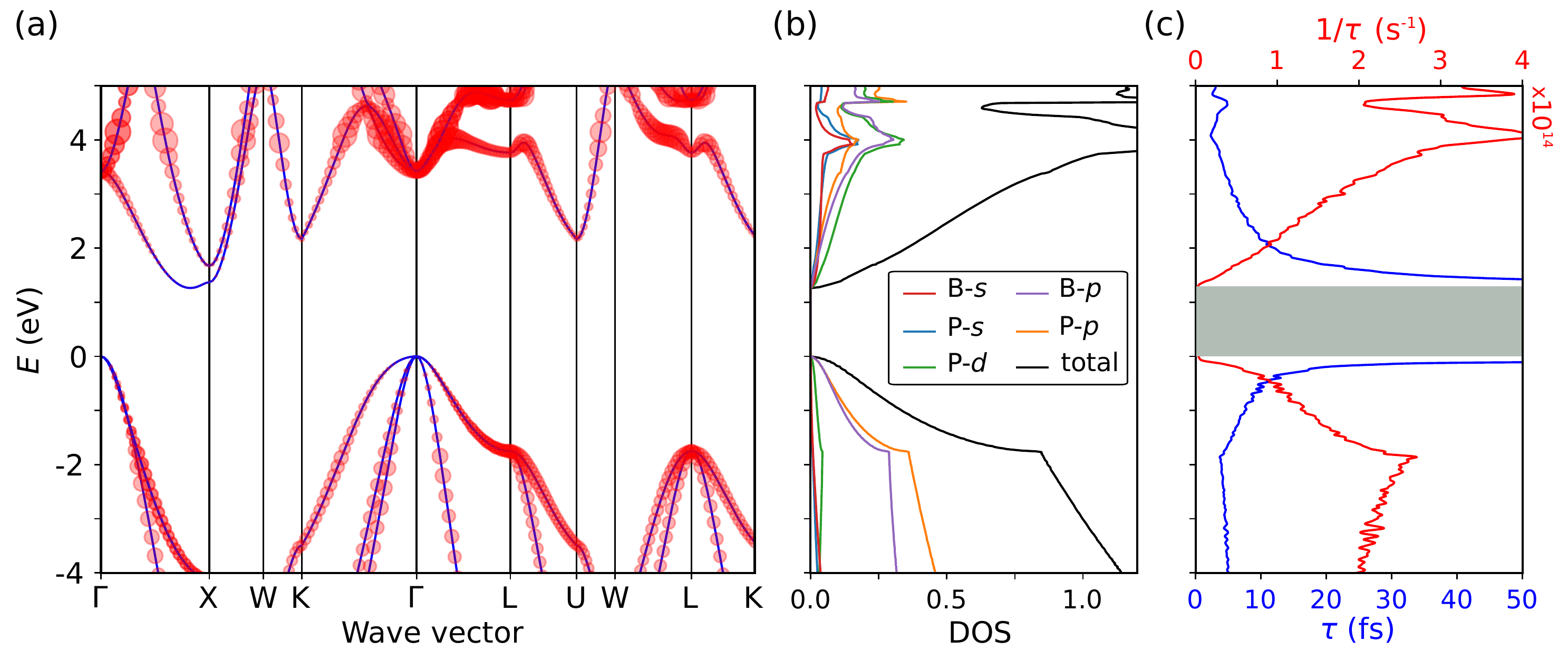}
\end{center}
\vspace{-15pt}
\caption{From left to right, (a) DFT electronic band structure at the PBE level, (b) projected DOS and (c) relaxation time $\tau$ and scattering rate $1/\tau$. Each DFT electronic state is marked with red dots representing its scattering rate by phonons at temperature of 300~K (the intensity of scattering is proportional to the size of the dots). The projected DOS are computed using DFT. The relaxation time $\tau$ (in femtoseconds) and its inverse, the scattering rate, $1/\tau$ (in 1/second) shown here as functions of the energy are calculated for undoped BP (the Fermi level is approximately set at its mid-gap).}
\label{scattering}
\end{figure*}

\begin{figure}[!ht]
\begin{center}
\includegraphics[width=0.9\linewidth]{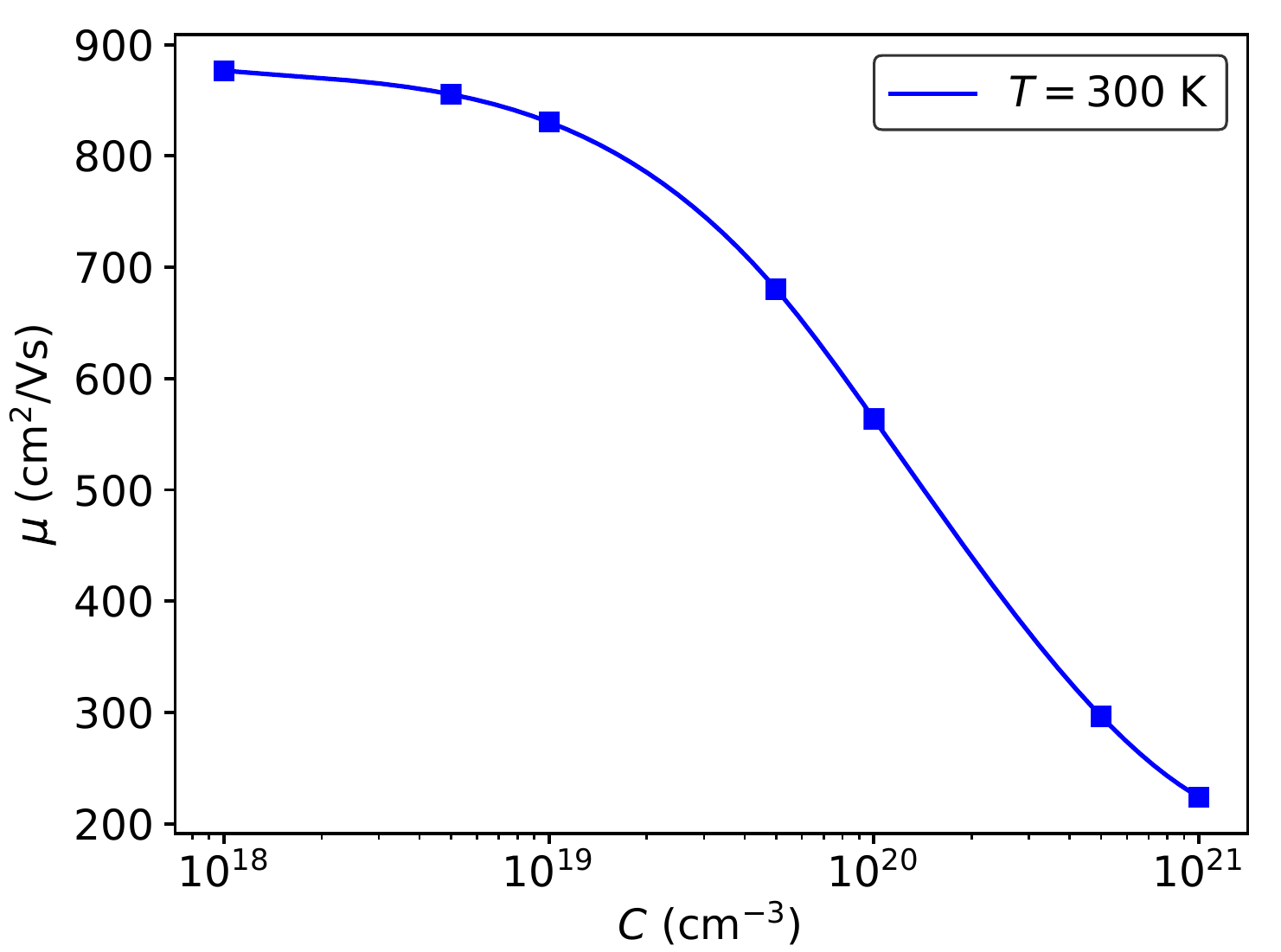}
\end{center}
\vspace{-15pt}
\caption{Hole mobilities computed as a function of hole concentration at 300~K. The curved line represents the interpolated data using spline interpolation method.}
\label{mobilities}
\end{figure}

Turning to electronic transport, \figref{scattering} (a) shows the DFT band structure and the scattering rates (inverse of $\tau_{n\mathbf{k}}$) at $300$~K for different electronic states calculated using the PBE functional. The radius of the red dots accounts for the intensity of the scattering rate of the corresponding electronic state. We also present the projected and total DOS (\figref{scattering} (b)) and the corresponding relaxation time and scattering rate (\figref{scattering} (c)) as functions of energy. In general, the scattering rates are proportional to the DOS, implying that the relaxation time becomes smaller in regions of high DOS. The Fermi level in doped BP can be located above or below the valence band maximum (VBM), depending on the number of holes and the density of states around this valley. For a hole concentration of $10^{18}$~cm$^{-3}$, the Fermi level is about 92~meV above the VBM, while it is located about 273~meV below the VBM for a very high concentration of $10^{21}$ cm$^{-3}$. We calculated scattering rates for different doping concentrations (see Fig. S2 in the supplementary document). From the band structure and the scattering rates, we can calculate the mobility at room temperature following Boltzmann transport theory. \figref{mobilities} plots the hole mobility in BP as a function of the hole concentration at $300$~K. The mobility decreases with the hole concentration because, as the Fermi level shifts below the VBM, the scattering rate increase (see \figref{scattering} (c) and Fig. S2 in the supplementary document). The hole mobility obtained at room temperature is about 900~cm$^2$/Vs at a doping concentration of $10^{18}$~cm$^{-3}$. This very high mobility is the signature of a very low scattering rate. Indeed, other materials with similar/lower effective masses show mobilities at similar carrier concentration and temperature that are significantly lower. For comparison, we provide the data (including calculated effective mass and mobility at carrier concentration of 10$^{18}$ cm$^{-3}$ and temperature 300~K) for some materials in Table SI of the supplementary document. \ce{Li3Sb}, which was also identified as a low hole effective mass material, shows hole effective masses of 0.24~\cite{Ricci2017, V.Ha2019} which are smaller than values of 0.34~\cite{J.B.Varley2017, Ricci2017} in BP. However, the phonon-limited hole mobility of \ce{Li3Sb} is only around 70 cm$^2$/Vs~\cite{V.Ha2019}. A similar tendency is also observed in oxides, e.g. the well-known high mobility \emph{n}-type TCO, \ce{BaSnO3}. This compound has an extremely low electron effective mass of 0.13~\cite{Hautier2014, Ricci2017} ($\sim$2.5 times smaller than the hole effective mass of BP), nonetheless, exhibits a calculated electron mobility (taking into account only longitudinal optical modes) of around 389 cm$^2$/Vs~\cite{Krishnaswamy2017}. This is $\sim$2.3 times smaller than the hole mobility of BP. 

The transport of holes in BP takes place around the $\Gamma$ point and has contributions from three bands. In this region of the Brillouin zone, the scattering is very weak (see \figref{scattering} (a)) leading to high relaxation times around the VBM (see \figref{scattering} (c) and Fig. S2 in the supplementary document). We hypothesize that this weak scattering comes from a very weak scattering from polar modes. Indeed, the Born effective charges~\cite{Gonze1997} computed for BP are very small~\cite{G.Petretto2018}: about 0.56 for B and -0.56 for P leading to weak long-range electron-phonon interactions~\cite{Verdi2015}. Both \ce{Li3Sb} and \ce{BaSnO3} are much more ionic. \ce{Li3Sb} shows Born effective charges between 0.71 and 1.46 for Li and around -2.89 for Sb. \ce{BaSnO3} Born effective charges are between 2.76 and 4.48 for the cations and between -3.51 and -1.86 for oxygen (see Table SI of the supplementary document). Both \ce{BaSnO3} and \ce{Li3Sb} should therefore show stronger scattering of electrons by polar phonons~\cite{Verdi2015}. 

On a more general note, this indicates that non-oxide compounds, in addition to lower hole effective masses, can show lower phonon scattering rates than oxides. This directly arises from the possibility for non-oxides being less ionic and offering therefore weaker polar phonon scattering of the electrons.

The computed mobility is an upper bound to experimental mobilities as it only takes into account intrinsic sources of scattering. We show in \tabref{table1} a series of experimental mobility measurements reported for boron phosphide. We note that some of the samples in these previous studies are not of very high quality, and the measured mobility depends on the morphology of fabricated samples, such as whether they are single crystals, polycrystalline, amorphous, or thin-film. Nevertheless, the experimental data confirms the potential for BP to deliver high mobilities.
\begin{center}
\setlength\tabcolsep{4pt}
\begin{table*}[!htb]
\caption{The fabrication methods, hole-carrier concentrations $C$ (in cm$^{-3}$) and mobilities $\mu$ (in cm$^2$/Vs) of cubic BP. These values are extracted from experimental measurements (as cited references) at room temperature. \label{table1}}
\renewcommand{\arraystretch}{1.2}
\begin{tabular*} {0.95\textwidth} {@{\extracolsep{\fill}} l l l l}
\hline
Processing      & $C$ (cm$^{-3}$)  & $\mu$ (cm$^2$/Vs) & Remarks \\
\hline
Solution (of Ni or Fe) growth & $1.0\times10^{18}$ & 500 \cite{C.C.Wang1964, T.L.Chu1971} & Contains 0.01\% solvent\\
Epitaxial growth & $8\times10^{19}$ & 285 \cite{K.Shohno1974} & Si substrate\\
Epitaxial growth & $5\times10^{19}$ & 350 \cite{K.Shohno1974} & Si substrate\\
Hetero-Epitaxial growth & $4.9\times10^{19}$ & 75.5 \cite{M.Takigawa1974} & Si substrate\\
Hetero-Epitaxial growth & $2.2\times10^{19}$ & 100.6 \cite{M.Takigawa1974} & Si substrate\\
Chemical transport technique & $1.67\times10^{18}$ & 1.77 \cite{Y.Kumashiro1984} & Small amount of \ce{B6P} mixed\\
CVD + thermal neutron irradiation & $1.1\times10^{17}$ & 100.3 \cite{Y.Kumashiro1988, Y.Kumashiro1990} & Si substrate\\
Photo-thermal CVD & $1.0\times10^{17}$ & 82 \cite{Y.Kumashiro2000} & Si substrate\\
\hline
\multicolumn{4}{ l }{CVD: Chemical Vapor Deposition}\\
\hline
\end{tabular*}
\end{table*}
\end{center}

Accordingly, our calculations show that boron phosphide offers a very high hole mobility ($\sim$900 cm$^2$/Vs) for a computed visible transmittance of 60\% in a 100-nm film. 
Note that this relatively low transmittance is mainly due to the reflectivity of BP, which accounts for 39\% of the loss of transparency. It is possible to use an antireflective coating to counter that problem~\cite{I.Hamberg1986, T.Fujibayashi2006}.
If a hypothetically perfect antireflective coating is used so that the reflectivity of BP is suppressed, the transmittance increases up to 98\% for a 100-nm film. 
This makes this material very attractive for transparent transistor applications. The possibility of \emph{n}- and \emph{p}-type doping of boron phosphide (as demonstrated both computationally and experimentally) strengthens even further its interest as a material for electronics applications in which ambipolar doping is a very useful property~\cite{Varley2013, Quackenbush2013}.

\begin{figure}[!htb]
\begin{center}
\includegraphics[width=0.9\linewidth]{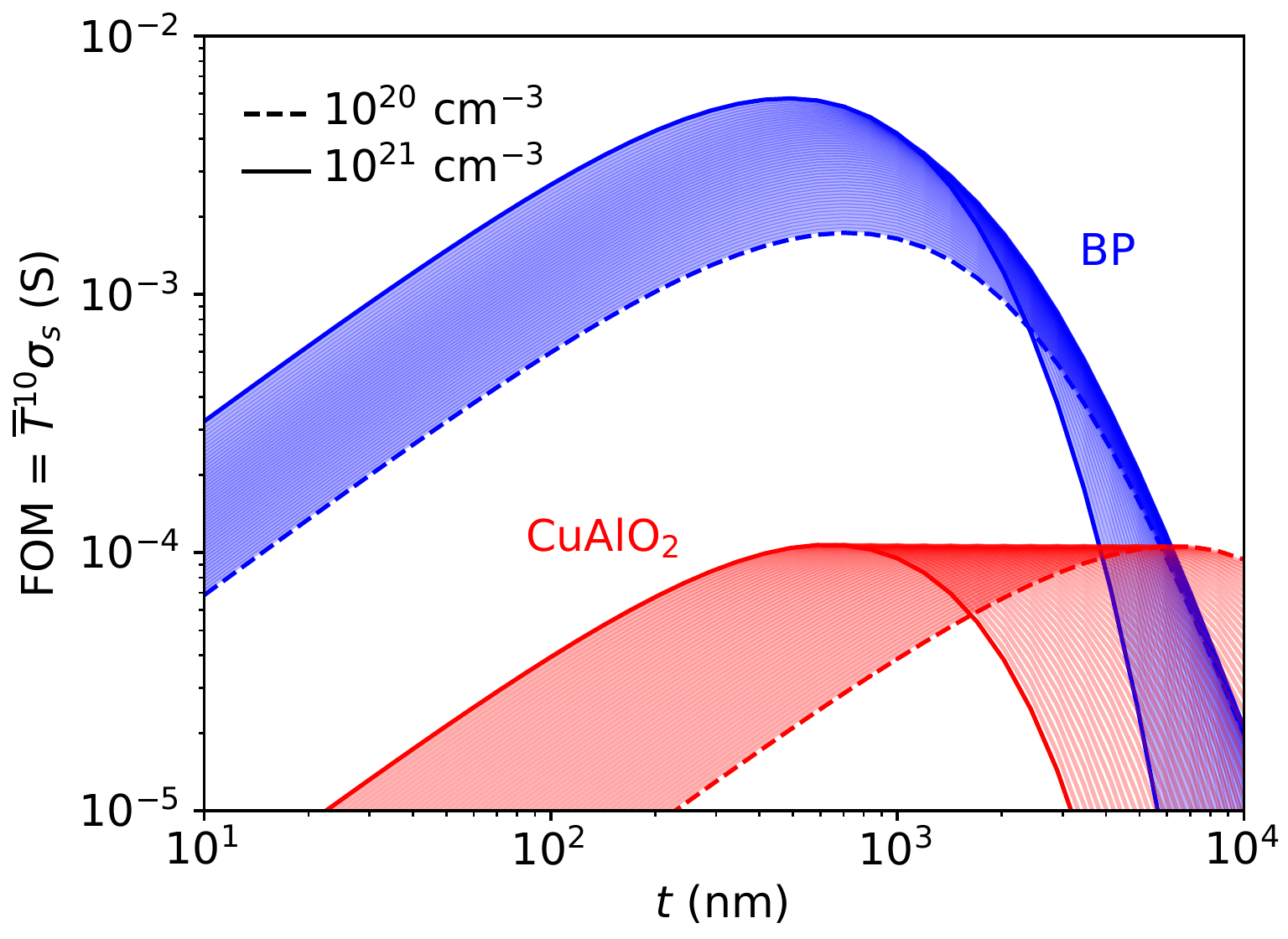}
\end{center}
\vspace{-15pt}
\caption{Haacke FOM of BP (blue) and \ce{CuAlO2} (red) as a function of the film thickness $t$. The dashed (solid) lines correspond to a hole concentration of $10^{20}$ ($10^{21}$) cm$^{-3}$.}
\label{figofmerit}
\end{figure}

In many applications such as contacts for solar cells, it is not the mobility but the conductivity that is the transport quantity of interest. In those applications, a compromise between transparency and conductivity is often looked for and FOMs have been constructed to evaluate material performance as TCMs~\cite{G.Brunin2019}. One of the most common FOM has been proposed by Haacke~\cite{Haacke1976}. It is obtained by multiplying the sheet conductivity by the transmittance to the power 10. FOMs are typically plotted versus thickness to find the optimal thickness. \figref{figofmerit} shows Haacke's FOM against thickness for one of the most traditional \emph{p}-type transparent oxide \ce{CuAlO2} (in red) versus boron phosphide (in blue). The different lines indicate the different doping concentrations from 10$^{20}$ to 10$^{21}$ cm$^{-3}$. Here, the hole mobilities corresponding to various hole densities are interpolated from the computed data as shown in \figref{mobilities}. Both materials show an optimal thickness around the $\mu$m length scale, but the achievable FOMs of BP are at least one order of magnitude larger. \ce{CuAlO2} has a fundamental band gap higher than 3 eV and therefore does not absorb light in the visible range. Despite BP absorbing in the visible, this only occurs through indirect transitions, and our analysis shows that the resulting low transparency is more than compensated by its higher conductivity in the overall FOM. Note that if reflectivity is suppressed in BP, its FOM increases by two orders of magnitude. Overall, our study confirms the interest of boron phosphide as a \emph{p}-type TCM and motivates further experimental work on this material and a reinvestigation of its growth and optoelectronic characterization~\cite{Fioretti2020}.

\section{Conclusions}
We have investigated the influence of phonons on the optical absorption and transport properties of BP using state-of-the-art \emph{ab initio} methods. We observe very weak phonon-assisted optical absorption in the visible range. The hole mobility is estimated by solving the Boltzmann transport equation taking into account phonon scattering. Our results show that BP has an exceptionally high hole mobility ($\sim$900 cm$^2$/Vs at low doping and room temperature) for a transparent material. This very high hole mobility is not only due to the low hole effective mass but also from the weak scattering of the electrons by phonons. This weak scattering is attributed to the weak ionic nature of boron phosphide leading to lower polar phonon scattering than in oxides. Our comparison of boron phosphide with established \emph{p}-type TCMs such as \ce{CuAlO2} indicates its much higher figure of merit and further confirms its potential as a \emph{p}-type TCM. On a more general note, as non-oxides tend to be less ionic than oxides, they can lead to lower polar phonon scattering and higher mobilities. This is one additional reason to explore further non-oxides for TCM applications.
%

\section{Acknowledgments}
V.-A.H. was funded through a grant from the FRIA. R.M. is grateful for financial support from MEXT-KAKENHI (17H05478 and 16KK0097), the FLAGSHIP2020 project (project nos. hp180206 and hp180175 at K-computer), the Toyota Motor Corporation, the I−O DATA Foundation, and the Air Force Office of Scientific Research (AFOSR-AOARD/FA2386-17-1-4049). G.-M.R. and G.B. are grateful to the F.R.S.-FNRS for financial support. B.M. acknowledges support from the Gianna Angelopoulos Programme for Science Technology and Innovation, and from the Winton Programme for the Physics of Sustainability. J.B.V. and G. H. acknowledge and thank Jim Edgar for useful discussions. We acknowledge access to various computational resources: the Tier-1 supercomputer of the F\'{e}d\'{e}ration Wallonie-Bruxelles funded by the Walloon Region (grant agreement N$^\circ$ 1117545); the facilities provided by the Universit\'{e} Catholique de Louvain (CISM/UCLouvain); the facilities provided by the Consortium des \'{E}quipements de Calcul Intensif en F\'{e}d\'{e}ration Wallonie Bruxelles (C\'{E}CI); the facilities of the Research Center for Advanced Computing Infrastructure at JAIST; and the Archer facility of the UK's national high-performance computing service for which access was obtained via the UKCP consortium [EP/P022596/1].
\bibliographystyle{apsrev4-1}
\bibliography{biblio}
\end{document}


\title{Supplemental Material for\\``Boron phosphide as a \emph{p}-type transparent conductor: optical absorption and transport through electron-phonon coupling''}
\author{Viet-Anh Ha}
\email[Present address: Oden Institute for Computational Engineering and Sciences, University of Texas at Austin, 201 E. 24$^{th}$ Street, Austin, TX 78712, USA]{}
 \affiliation{Institute of Condensed Matter and Nanoscience (IMCN), Universit\'{e} Catholique de Louvain (UCL), Chemin \'{e}toiles 8, bte L7.03.01, Louvain-la-Neuve 1348, Belgium}
\author{Bora Karasulu}%
 \affiliation{Cavendish Laboratory, University of Cambridge, J. J. Thomson Avenue, Cambridge CB3 0HE, United Kingdom}%
 \author{Ryo Maezono}%
 \affiliation{School of Information Science, JAIST, Asahidai 1-1, Nomi, Ishikawa 923-1292, Japan}%
 \author{Guillaume Brunin}
 \affiliation{Institute of Condensed Matter and Nanoscience (IMCN), Universit\'{e} Catholique de Louvain (UCL), Chemin \'{e}toiles 8, bte L7.03.01, Louvain-la-Neuve 1348, Belgium}
 \author{Joel Basile Varley}%
 \affiliation{Lawrence Livermore National Laboratory 7000 East Avenue, L-413 Livermore, CA 94550, USA}%
\author{Gian-Marco Rignanese}%
 \affiliation{Institute of Condensed Matter and Nanoscience (IMCN), Universit\'{e} Catholique de Louvain (UCL), Chemin \'{e}toiles 8, bte L7.03.01, Louvain-la-Neuve 1348, Belgium}%
\author{Bartomeu Monserrat}%
 \email[\emph{E-mail}: ]{bm418@cam.ac.uk}
 \affiliation{Cavendish Laboratory, University of Cambridge, J. J. Thomson Avenue, Cambridge CB3 0HE, United Kingdom}
 \affiliation{Department of Materials Science and Metallurgy, University of Cambridge, 27 Charles Babbage Road, Cambridge CB3 0FS, United Kingdom}%
\author{Geoffroy Hautier}%
 \email[\emph{E-mail}: ]{geoffroy.hautier@uclouvain.be}
 \affiliation{Institute of Condensed Matter and Nanoscience (IMCN), Universit\'{e} Catholique de Louvain (UCL), Chemin \'{e}toiles 8, bte L7.03.01, Louvain-la-Neuve 1348, Belgium}%
\date{\today}
\maketitle
%
We use electronic and phonon band structures computed using the hybrid HSE functional for phonon-assisted optical absorption in order to better reproduce the magnitude of the band gap, while we use electronic and phonon band structures computed using the semilocal PBE functional in the transport calculations. \figref{phband-hse} compares the phonon band structures computed by the PBE and HSE functionals, with the HSE functional giving slightly higher vibrational frequencies, particularly for optical modes.
\begin{myfigure}[htbp]
\begin{center}
\includegraphics[width=0.9\linewidth]{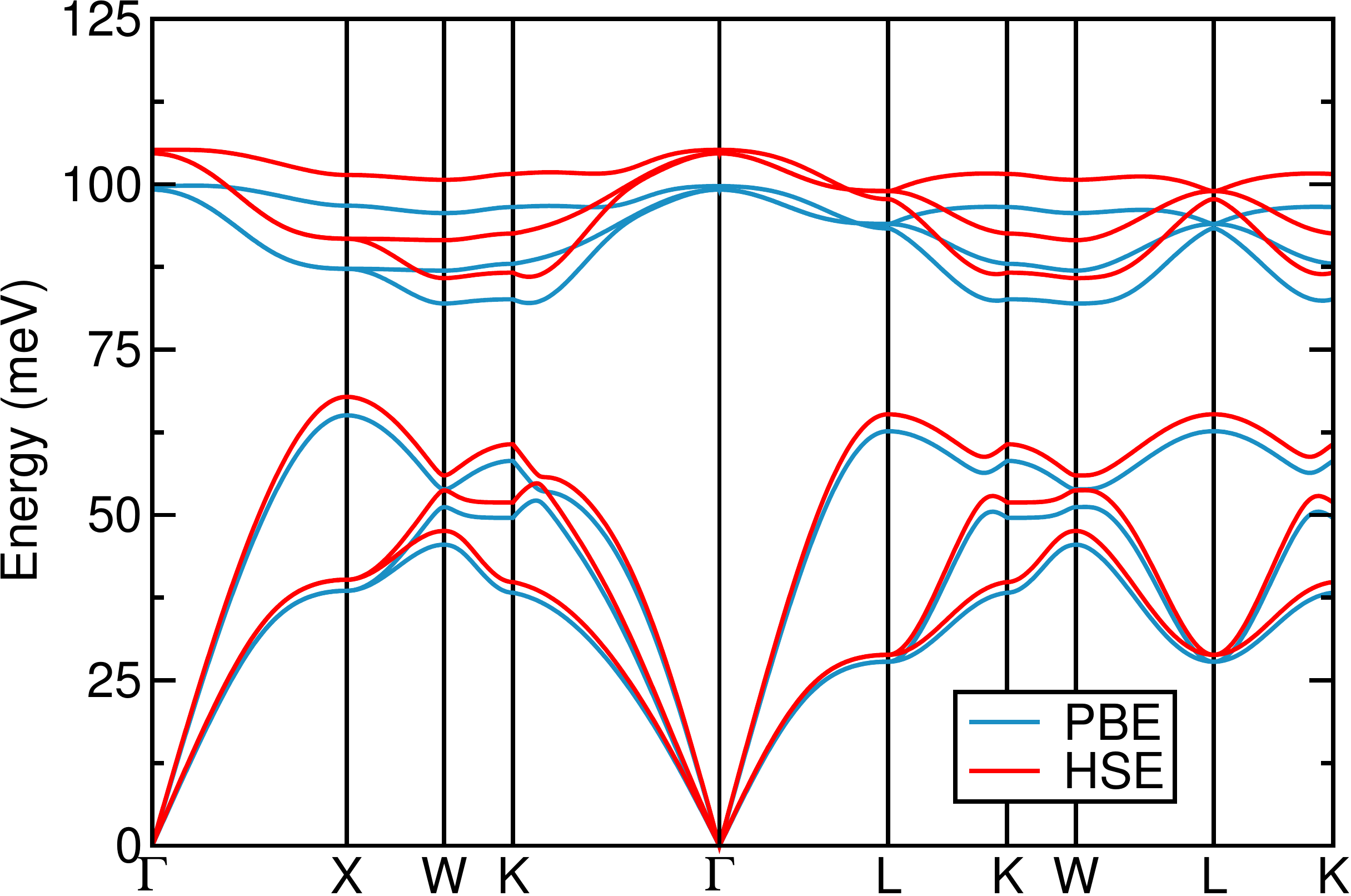}
\end{center}
\vspace{-15pt}
\caption{Phonon band structures comparing the PBE and the HSE functionals. The HSE functional leads to higher vibrational energies.}
\label{phband-hse}
\end{myfigure}

\figref{scattering} plots scattering rates as functions of energy for different doping concentrations such as $10^{18}$, $5\times10^{18}$, $10^{19}$, $5\times10^{19}$, $10^{20}$, $5\times10^{20}$, and $10^{21}$ cm$^{-3}$. The Fermi level corresponding to each doping concentration is shown. The scattering does not change significantly in advance of heavily-doped regimes since $5\times10^{20}$ cm$^{-3}$ in which the Fermi levels move very deeply into the valence band.
\begin{myfigure}[htbp]
\begin{center}
\includegraphics[width=0.9\linewidth]{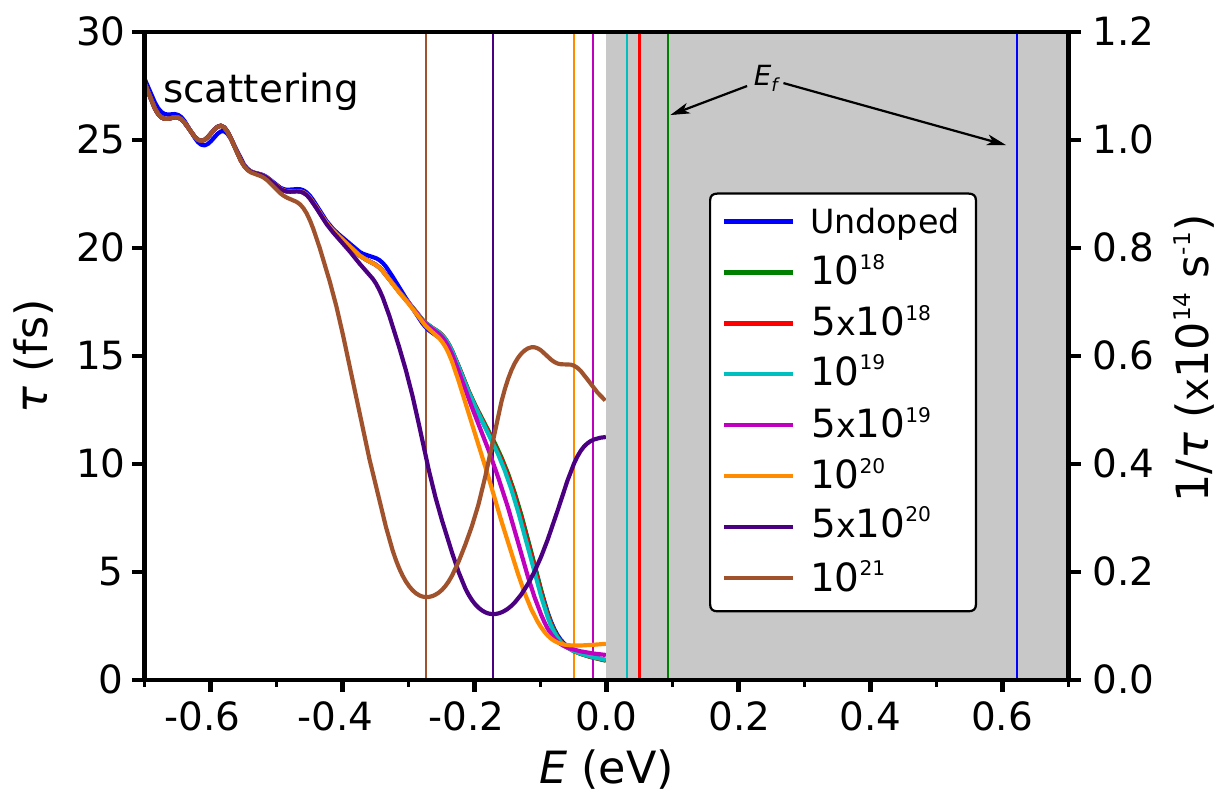}
\end{center}
\vspace{-15pt}
\caption{Scattering rates for different hole concentrations. The corresponding Fermi levels are indicated by vertical lines. Valence band maximum is set to zero.}
\label{scattering}
\end{myfigure}

\begin{mytable}[!h]
\caption{Formula, there eigenvalues of (hole/electron) effective mass tensor (in unit of free-electron mass $m_o$), three eigenvalues of Born effective charge tensor of each atom in the unit cell (in unit of the elementary charge $e$) and average calculated (hole/electron) mobility (in cm$^2$/Vs) at carrier concentration of 10$^{18}$ (cm$^{-3}$) and room-temperature 300~K. The corresponding references are marked.}
\label{mbltvsefftchg}
\begin{center}
\renewcommand{\arraystretch}{1.3}
\begin{tabular}{@{\extracolsep{\fill}}|l|l|l|l|}
\hline
Formula & Effective mass~\cite{Ricci2017} & Born effective charge~\cite{G.Petretto2018} & Mobility \\
\hline
\multicolumn{1}{ |l  }{\multirow{2}{*}{BP} } &
\multicolumn{1}{ |l| }{\multirow{2}{*}{$m_h = [0.34,\ 0.34,\ 0.34]$} } & B: $[-0.558,\ -0.558,\ -0.558]$ & \multicolumn{1}{ |l| }{\multirow{2}{*}{$\mu_h \sim 900$~[This work]} }     \\
\multicolumn{1}{ |l  }{} &
\multicolumn{1}{ |l| }{} & P: $[0.558,\ 0.558,\ 0.558]$ & \multicolumn{1}{ |l| }{}     \\ \cline{1-4}
\multicolumn{1}{ |l  }{\multirow{2}{*}{CaTe} } &
\multicolumn{1}{ |l| }{\multirow{2}{*}{$m_h = [0.60,\ 0.60,\ 0.60]$} } & Ca: $[2.381,\ 2.381,\ 2.381]$ & \multicolumn{1}{ |l| }{\multirow{2}{*}{$\mu_h \sim 20$~\cite{V.Ha2019}} }     \\
\multicolumn{1}{ |l  }{} &
\multicolumn{1}{ |l| }{} & Te: $[-2.381,\ -2.381,\ -2.381]$ & \multicolumn{1}{ |l| }{}     \\ \cline{1-4}
\multicolumn{1}{ |l  }{\multirow{4}{*}{\ce{Li3Sb}} } &
\multicolumn{1}{ |l| }{\multirow{4}{*}{$m_h = [0.24,\ 0.24,\ 0.24]$} } & Li-1: $[1.460,\ 1.460,\ 1.460]$ & \multicolumn{1}{ |l| }{\multirow{4}{*}{$\mu_h \sim 70$~\cite{V.Ha2019}} }     \\
\multicolumn{1}{ |l  }{} &
\multicolumn{1}{ |l| }{} & Li-2: $[0.713,\ 0.713,\ 0.713]$ & \multicolumn{1}{ |l| }{}     \\
\multicolumn{1}{ |l  }{} &
\multicolumn{1}{ |l| }{} & Li-3: $[0.713,\ 0.713,\ 0.713]$ & \multicolumn{1}{ |l| }{}     \\
\multicolumn{1}{ |l  }{} &
\multicolumn{1}{ |l| }{} & Sb: $[-2.886,\ -2.886,\ -2.886]$ & \multicolumn{1}{ |l| }{}     \\ \cline{1-4}
\multicolumn{1}{ |l  }{\multirow{5}{*}{\ce{BaSnO3}} } &
\multicolumn{1}{ |l| }{\multirow{5}{*}{$m_e = [0.13,\ 0.13,\ 0.13]$} } & Ba: $[2.755,\ 2.755,\ 2.755]$ & \multicolumn{1}{ |l| }{\multirow{5}{*}{$\mu_e \sim 389$~\cite{Krishnaswamy2017}} }     \\
\multicolumn{1}{ |l  }{} &
\multicolumn{1}{ |l| }{} & Sn: $[4.482,\ 4.482,\ 4.482]$ & \multicolumn{1}{ |l| }{}     \\
\multicolumn{1}{ |l  }{} &
\multicolumn{1}{ |l| }{} & O-1: $[-1.862,\ -1.862,\ -3.513]$ & \multicolumn{1}{ |l| }{}     \\
\multicolumn{1}{ |l  }{} &
\multicolumn{1}{ |l| }{} & O-2: $[-1.862,\ -3.513,\ -1.862]$ & \multicolumn{1}{ |l| }{}     \\ 
\multicolumn{1}{ |l  }{} &
\multicolumn{1}{ |l| }{} & O-3: $[-3.513,\ -1.862,\ -1.862]$ & \multicolumn{1}{ |l| }{}     \\ \cline{1-4}
\end{tabular}
\end{center}
\end{mytable}

\tabref{mbltvsefftchg} presents information about (hole/electron) effective mass, effective charge and calculated (hole/electron) mobility for several materials. The effective masses and the effective charges are referred from Refs.~\cite{Ricci2017} and~\cite{G.Petretto2018}, respectively. We note that the effective masses here are the average quantities which take into account the contribution from all bands at a given doping concentration and temperature (the effective masses shown here correspond to a carrier concentration of 10$^{18}$ cm$^{-3}$ and temperature of $300$~K). More details about the definition and computational method can be found in Ref.~\cite{Hautier2014}. The electron mobility of \ce{BaSnO3} in \tabref{mbltvsefftchg} was reported by Krishnaswam and co-workers~\cite{Krishnaswamy2017}, which is estimated with only taking into account the influence of longitudinal optical phonon-modes.
%

%
%
%
%
\FloatBarrier
\bibliography{supbiblio}